\documentclass[prd, twocolumn, nofootinbib, floatfix]{revtex4}

\usepackage{epsfig} 
\usepackage{amsmath}

\newcommand{\beq}{\begin{equation}}
\newcommand{\eeq}{\end{equation}}
\newcommand{\beqa}{\begin{eqnarray}}
\newcommand{\eeqa}{\end{eqnarray}}
\newcommand{\om}{\Omega_m}

\newcommand{\ls}{\mathrel{\raise0.27ex\hbox{$<$}\kern-0.70em \lower0.71ex\hbox{{
$\scriptstyle \sim$}}}}

\begin{document} 

\title{Exponential Gravity} 
\author{Eric V.\ Linder} 
\affiliation{Berkeley Lab \& University 
of California, Berkeley, CA 94720, USA \\ 
Institute for the Early Universe, Ewha Womans University, Seoul, South Korea} 
\date{\today}

\begin{abstract} 
We investigate a $f(R)$ modification of gravity that is exponential in the 
Ricci scalar $R$ to explain cosmic acceleration.  The steepness of 
this dependence provides extra 
freedom to satisfy solar system and other curvature regime constraints. 
With a parameter to alleviate the usual fine tuning of 
having the modification strengthen near the present, the total number 
of parameters is only one more than $\Lambda$CDM.  The resulting 
class of solutions asymptotes to $w=-1$ but has no cosmological constant. 
We calculate the dynamics in detail, examine the effect on the matter 
power spectrum, and provide a simple fitting form relating the two. 
\end{abstract} 

\maketitle

\section{Introduction \label{sec:intro}}

Einstein's general relativity has proved to be a remarkably 
robust theory of gravity.  Large numbers of attempts have been made 
to modify it, e.g.\ for the purpose of explaining cosmic acceleration 
without adding a separate dark energy component.  Many of these have 
been found to have pathologies such as ghosts, unbounded energies, 
tachyons, or other ills \cite{durrer}.  Gravity theories that are 
sound may still 
fail to provide acceleration, an early matter dominated period, or 
fall afoul of gravity constraints on solar system or other scales. 

Here we work within the framework of $f(R)$ theories that generalize 
the linear dependence on the Ricci scalar $R$ in the Einstein-Hilbert 
action.  Many such 
viable models exist but tend to have a restricted range obeying 
structure constraints, and in this regime are effectively identical 
to $\Lambda$CDM as far as the expansion history and distances.  
Furthermore these often take the characteristic curvature scale 
for the modification to the action to be that of the present 
matter density, i.e.\ when $R=8\pi G\rho_{\rm today}$, so that the 
modifications kick in near today.  

We explore ways of ameliorating all three issues: easing the 
restrictions of solar system tests and structure constraints on 
the curvature evolution, loosening the expansion history, and 
relaxing the fine tuning. 

Considerable work has been done on $f(R)$ theories; see \cite{sotiriou,araa} 
for overall reviews and references.  In this paper we follow most closely 
the setup of \cite{husaw,chibasmith,songhusaw}.  Our aim is to investigate 
the dynamics in more detail than usual, and relate it to the growth effects 
more explicitly, keeping close touch with the cosmological observations of 
both.  In \S\ref{sec:expan} we 
describe the model, its equations of motion, and the solutions for the 
expansion history and effective equation of state.  The effects on 
systems with high curvature and density gradients, such as the solar 
system and galaxies, are discussed in \S\ref{sec:solsys} along with 
the growth of structure.

\section{Model and Dynamics \label{sec:expan}} 

The $f(R)$ class of extensions to general relativity represents a 
simple modification that preserves the maximum number of derivatives 
entering at two, and can have a well defined limit in which the 
theory reduces to general relativity, typically in the high scalar 
curvature regime.  The gravitational part of the modified 
Einstein-Hilbert action is 
\beq 
S=\frac{1}{16\pi G}\int d^4x\,\sqrt{-g}\,\,[R+f(R)]\,, 
\eeq 
where $g$ is the determinant of the metric and $R$ is the Ricci 
scalar curvature.  

In order to avoid including an implicit cosmological 
constant we require the low curvature limit of $f$ to vanish, 
i.e.\ $f(0)=0$.  At high curvature (e.g.\ in the early universe), we want 
the modification term 
to be negligible compared to the usual term linear in $R$, so we require 
$f(R\gg1)/R\to0$.  We do not want to put in by hand a specific 
preference for dynamics at the present, i.e.\ a fine tuning, so we 
do not set the characteristic scale of $f$ in terms of 
$R_0=8\pi G\rho(a=1)$, where $a$ is the cosmic scale factor and 
$\rho$ the matter density.  Rather, 
we give $f$ a transition scale $r$ and will fit for $r$ given 
observations.  Finally, in order to satisfy the constraints from 
locally high curvature systems such as the solar system or galaxies 
that gravity must be very close to general relativity, we take $f$ 
to have a steep dependence on scalar curvature, rapidly restoring 
Einstein gravity.  

This steepness will be a key ingredient in 
improving agreement with structure constraints while ameliorating 
fine tuning.  Steep potentials have been considered previously 
(see, e.g., \cite{zhang,husaw,appbat,tsuji,odintsov}) but here 
we explore in substantial detail the dynamics, from the deviation from 
matter domination, to the maximum equation of state variation, to the 
long time behavior in the equation of state function $w(a)$ and phase 
space $w$-$w'$.  We also give a quantitative analysis of effects on the 
matter power spectrum, as well as an accurate fitting form relating the 
expansion and growth effects.  These results are of interest apart from 
the specific form adopted.  
We now seek the simplest form with fewest parameters yet 
freedom to fit over a variety of characteristic curvature scales, 
not putting in special behavior just at the present density. 

All these conditions are satisfied by the ansatz 
\beq 
f(R)=-c\,r\,(1-e^{-R/r})\,. \label{eq:fdef} 
\eeq 
This simple form involves two quantities, and in fact we will see that 
the combination $cr$ is equivalent to setting the dimensionless matter 
density today $\om$, so there is only one parameter besides $\om$.  
Thus for this form the complexity is just one step more than the standard 
$\Lambda$CDM model. 
We focus on cosmological 
gravity from the matter domination through the accelerating eras, i.e.\ 
from high curvature to the asymptotic future; note primordial 
nucleosynthesis will not be altered, as discussed in Sec.~\ref{sec:solsys}. 

The Friedmann equation of motion is modified to \cite{songhusaw} 
\beqa  
H^2+\frac{f}{6}&-&f_R\,\left[\frac{1}{2}(H^2)'+H^2\right]+H^2 f_{RR}R' 
\nonumber \\ 
&=&\frac{8\pi G}{3}\rho(a)\equiv m^2 a^{-3}\,, \label{eq:heom}
\eeqa 
where a prime denotes $d/d\ln a$, a subscript $R$ denotes a derivative 
with respect to $R$, $H$ is the Hubble parameter, and $\rho$ is the 
matter density, taken to be the only physical component of energy density. 
Note $R=12H^2+3(H^2)'$, and for our exponential gravity model 
$f_{RR}=(c/r)\,e^{-R/r}$. 
Thus $f_{RR}>0$, a critical stability condition (see \cite{araa} for 
a review). 

Following \cite{husaw} with some modifications we define 
\beqa 
x_H&=&\frac{H^2}{m^2}-a^{-3}-\frac{c}{6}\frac{r}{m^2}\,, \\ 
x_R&=&\frac{R}{m^2}-3a^{-3}-2c\frac{r}{m^2}-12x_H\,, 
\eeqa 
to take out the leading order terms.  The equation of motion (\ref{eq:heom}) 
becomes two coupled first order equations 
\beqa 
x_H'&=&x_R/3 \,, \\ 
x_R'&=&-4x_R-\frac{1}{m^2f_{RR}} 
\frac{x_H+[d-9(cm^2/r)a^{-6}]e^{-R/r}}{a^{-3}+x_H+d} \nonumber \\ 
&+&9\,\frac{x_H+d}{1+a^3(x_H+d)}+\frac{f_R}{m^2 f_{RR}} 
\left(\frac{R}{6H^2}-1\right)\,, 
\eeqa 
where $d=cr/(6m^2)$ and 
\beq 
\frac{R}{6H^2}=\frac{1}{2}\,\frac{1+a^3(4d+4x_H+x_R/3)}{1+a^3(x_H+d)}\,. 
\eeq 
In this form we have carried out the explicit cancellation of several 
terms that nominally appeared dominant, making the numerical solution more 
robust. 

Treating the terms modifying the Friedmann equation (\ref{eq:heom}) 
as an effective 
dark energy density $\Omega_{de}(a)$, we can define the effective dark 
energy equation of state and its variation, 
\beqa 
\Omega_{de}(a)&=&\frac{x_H+d}{a^{-3}+x_H+d}\,, \label{eq:oma} \\ 
w&=&-1-\frac{1}{9}\,\frac{x_R}{x_H+d}\,, \label{eq:wx}\\ 
w'&=&3\,(1+w)^2+(1+w)\frac{x_R'}{x_R}\,. \label{eq:wpw} 
\eeqa 

In the future, we see that $\Omega_{de}\to1$, 
i.e.\ the modifications dominate and the effective dark energy density 
goes to the critical density.  
In the high curvature regime, $f$ goes to a constant, $-cr$, and so 
one expects this to be related to $\Omega_\Lambda$, or $1-\Omega_m$ in 
the spatially flat universe we consider.  In general, the exact 
expression for $cr$ is given by 
\beq 
d\equiv\frac{c}{6}\frac{r}{m^2}=\om^{-1}-1-x_H \,, \label{eq:dcr}
\eeq 
so there is a defined relation between $cr$ and $\om$.  When the 
curvature is large, $x_H\sim e^{-R/r}$ (see below) and can be neglected 
and the product $cr$ is 
explicit in terms of $\om$.  Since $cr$ corresponds to $\om$ 
(all plots are for $\om=0.28$), this 
leaves $c$ as the only other parameter of the theory. 

The dynamics can also be understood fairly simply.  In the high 
curvature limit when both $x_H$ and $x_R$ are small, $w\to-1$, i.e.\ 
it starts from a frozen, cosmological constant state during the 
matter dominated expansion.  Expanding the equations of motion at 
high redshift one finds that $x_H$, $x_R\sim e^{-R/r}$ and are positive.  
So from Eq.~(\ref{eq:wx}) one sees that the field starts off phantom, 
i.e.\ with $w<-1$, and since $x_R'/x_R~\sim a^{-3}$ then from 
Eq.~(\ref{eq:wpw}) 
the field evolves quite rapidly, $w'\sim a^{-3}(1+w)$, much faster than 
a physical thawing scalar field which has $w'=3(1+w)$ \cite{caldlin,cahndl}. 

\begin{figure}[!htb]
\begin{center}
\psfig{file=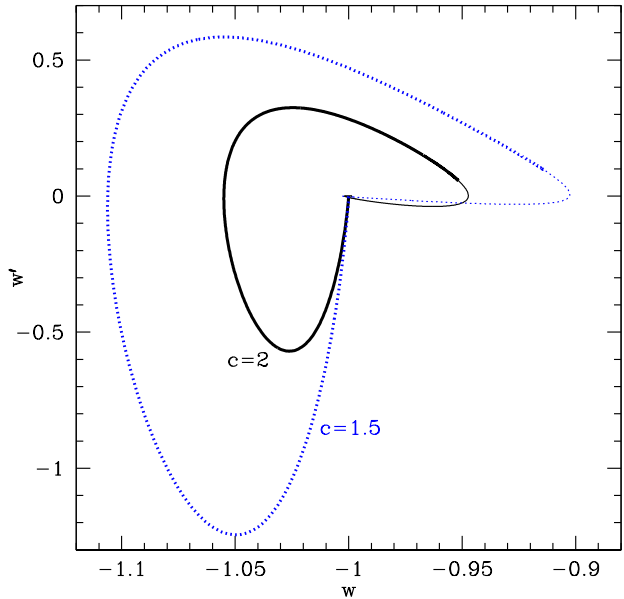,width=3.4in}
\caption{The phase space dynamics of the effective dark energy 
parameters $w'$-$w$ is unlike any standard scalar field.  It crosses 
between the phantom and normal parts of the phase space (often a signal 
of modified gravity), and ``sublimates'' rather than thaws from 
the initial frozen state.  Successive oscillations about $\Lambda$ 
are barely visible here, showing rapid damping to the asymptotic 
cosmological constant.  Thick portions of the curves show the 
evolution for $a\le1$. 
}
\label{fig:wwp}
\end{center}
\end{figure}

Solving the equations of motion numerically, we display the effective 
phase space dynamics of $w'$-$w$ in Fig.~\ref{fig:wwp}.  We indeed see 
evolution from a frozen, cosmological constant state ($w=-1$, $w'=0$) 
to the phantom regime, more swiftly than thawing (``sublimation''), 
and then an oscillatory behavior where the field reaches a maximum 
value of $|1+w|$, then crosses back over $w=-1$, reaches a secondary 
maximum and quickly damps around $w=-1$.  Note the second period of 
oscillations is highly damped, so small on the figure it appears as 
just a short dash around $w=-1$.

In fact, there is an asymptotic behavior toward an effective cosmological 
constant.  The stationary point of the equation of motion is given by 
\beqa 
x_R&=&0\,, \\ 
x_H&=&-d\,(c+1)\,(e^{R/r}+c)^{-1}\,, 
\eeqa 
with an implicit relation for $R$ of 
\beq 
\frac{R}{m^2}=12d\,\frac{e^{R/r}-1}{e^{R/r}+c}\,. \label{eq:rfut} 
\eeq 
So $R$ settles asymptotically to a constant value, as does $H$, meaning 
that the future solution possesses $w=-1$.  When the asymptotic value of 
$e^{R/r}$ is much greater than one then $R/m^2\to 12d$, or 30.86 for 
$\om=0.28$.  No 
valid attractor solution exists for $c<1$, because this would require $R<0$. 

\begin{figure}[!htb]
\begin{center}
\psfig{file=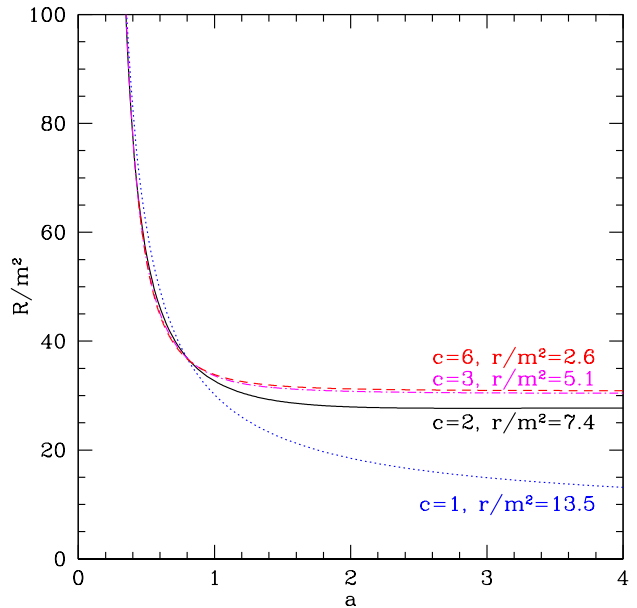,width=3.4in}
\caption{The Ricci scalar curvature, normalized by the present matter 
density, evolves very rapidly from the high redshift, high curvature 
state to a constant asymptotic state (for $c>1$).  As $c$ gets larger 
than unity, $R/m^2$ approaches a constant given by $12\,(\om^{-1}-1)$ 
(i.e.\ 30.86 for $\om=0.28$ as here) and the modification factor $e^{R/r}$ 
always stays large.  The influence 
of the modification occurs near the 
present for any allowed values of $r$, without additional fine tuning. 
}
\label{fig:ric}
\end{center}
\end{figure}

Figure~\ref{fig:ric} illustrates the numerical solution for $R(a)$ 
for several values of $c$.  We see that as promised $R$ evolves 
very steeply due to the exponential form of $f(R)$ and is in the 
high curvature limit $e^{R/r}\gg1$ until recently.  It then quickly 
reaches its asymptotic value given by Eq.~(\ref{eq:rfut}).  As $c$ 
increases, $r$ necessarily decreases to preserve $cr$, i.e.\ $\om$, 
and so $e^{R/r}$ is always large and asymptotically $R/m^2=30.86$.  
Thus, the evolution of $R(a)$ is quite similar for all $c\gtrsim3$ 
($r/m^2\lesssim5$). 
The influence of the modification rises near the 
present for any allowed values of $r$, without additional fine tuning.

\begin{figure}[!htb]
\begin{center}
\psfig{file=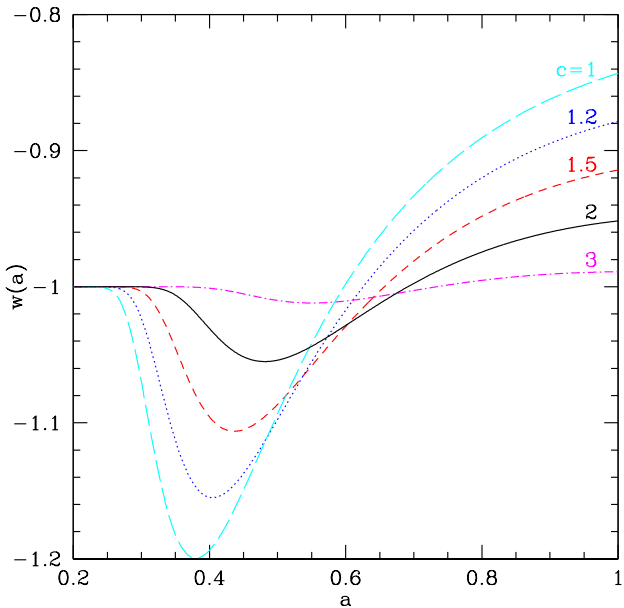,width=3.4in}
\caption{The effective dark energy equation of state evolution is shown 
for various values of $c$.  As $c$ gets large, the expansion history 
becomes indistinguishable from $\Lambda$CDM. 
}
\label{fig:wofa}
\end{center}
\end{figure}

\begin{figure}[!htb]
\begin{center}
\psfig{file=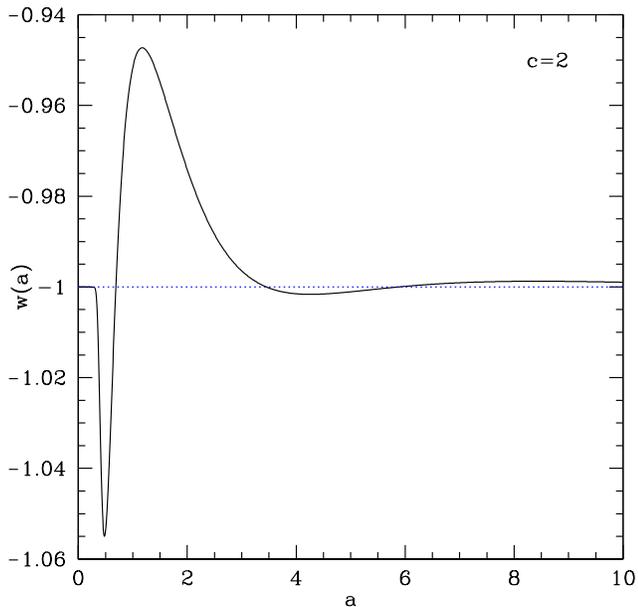,width=3.4in}
\caption{The long term history of the effective dark energy equation 
of state is evolution from a cosmological constant state ($w=-1$), 
deviation to both the phantom ($w<-1$) and normal ($w>-1$) sides, and 
rapid damping to a future cosmological constant state. 
}
\label{fig:walong}
\end{center}
\end{figure}

To examine the expansion history in more detail, we consider the 
effective dark energy equation of state $w(a)$ induced by the 
gravity modifications.  Figure~\ref{fig:wofa} shows this function for 
a variety of different values of $c$.  As discussed, the equation of 
state goes from $w=-1$ at high redshift to $w<-1$ then crosses back 
to $w>-1$.  The maximum departure from $w=-1$ decreases rapidly as 
$c$ increases, going roughly as $e^{-2c}$ for large $c$.  
Thus as $c$ gets very large, any deviations from $\Lambda$CDM are 
strongly suppressed.  
Figure~\ref{fig:walong} extends the plot to the future, showing the 
rapidly damped oscillation around $w=-1$.

From the perspective of 
expansion history measurements, the distance to the cosmic microwave 
background last scattering surface agrees with the $\Lambda$CDM model 
with the same present matter density to 0.54\% (0.2\%) for all valid 
$c$ (for $c>1.5$).  
Distances to redshifts $z<2$, e.g.\ as measured by the Type Ia 
supernovae magnitude-redshift relation, agree to 2.3\% (1\%) for 
all valid $c$ (for $c>1.6$).  The parenthetical values for each 
correspond roughly 
to next generation observational limits.  However, we will see in the next 
section that cosmic structure and its growth impose more severe 
limits on the parameter values allowed.

\section{Curvature, Structure, and Growth \label{sec:solsys}} 

The $f(R)$ class of models of most interest acts like general 
relativity at high curvatures, so as to preserve agreements with 
primordial nucleosynthesis and other early time physics.  However 
they possess additional scalar degrees of freedom that can become 
apparent in lower curvature or high density gradient regimes 
\cite{husaw}.  Treating $f_R$ as a scalar field one can define 
a Compton wavelength 
below which the effects on structure formation and bound structures 
become significantly different from general relativity.  For example, 
the first parametrized post-Newtonian parameter, $\gamma$, takes the 
value $1/2$ rather than unity as in general relativity.  (In the 
high curvature regime inside the structure, general relativity is 
restored by the chameleon mechanism \cite{khoury}, under conditions 
detailed by, e.g., \cite{chibasmith}, which the exponential model 
can satisfy.  We address this further below.)  

The key indicators to these effects 
are the quantities $f_R$ and the effective Compton wavelength with 
respect to $f_R$, usually written in terms of \cite{songhusaw,husaw} 
\beq 
B=\frac{f_{RR}}{1+f_R}\,R'\,\frac{2H^2}{(H^2)'}\,, 
\eeq 
where $B$ is the square of the wavelength in units of the Hubble scale. 
Note that $1+f_R>0$ in the exponential model for all $c$ giving 
a finite attractor solution, 
i.e.\ $c>1$.  This even holds during the radiation era since although 
we are used to thinking that $R=0$ then, the value is in fact 
$R\ll\rho_{\rm rad}$ -- but one still has 
$R\sim 8\pi G\rho_{m,0}a^{-3}\gg 8\pi G\rho_{m,0}\sim r$ so $R/r\gg1$ 
at least back through primordial nucleosynthesis and so $1+f_R\to1$.  
The condition $B>0$, together with the vanishing of derivatives of $f$ 
at high redshift, delivers a standard radiation era \cite{songhusaw}.  
Given the scale $B$, modifications to structure occur for 
wavenumbers 
$k> aH/B^{1/2}$.  This will begin to affect linear perturbation 
growth, $k<0.1\,h$/Mpc say, for $B\gtrsim 10^{-5}$.  We will later 
calculate a more exact observational constraint. 

The exponential form of Eq.~(\ref{eq:fdef}) ameliorates the issues 
involved with deviations from general relativity in structure 
constraints, since $f_R$ and 
$B\sim e^{-R/r}$.  First, the evolution of the 
curvature and its effects are much more rapid than in many $f(R)$ theories, 
so for a given present value of curvature, or $f_R$, the deviation from 
general relativity can be much smaller in the past, when the 
structure formed, at $a=0.5$ say.  To rigorously quantify this 
argument requires numerical simulations of 
nonlinear structure formation in the specific model of modified gravity to 
robustly compute the ``leaking'' of the non-general relativity influence 
into the structure 
over time, as in \cite{simul}.  
If we can take advantage of this increased latitude, we can 
allow the characteristic curvature scale $r$ to be more relaxed from the 
present matter density, giving greater freedom in the model. 

\begin{figure}[!htb]
\begin{center}
\psfig{file=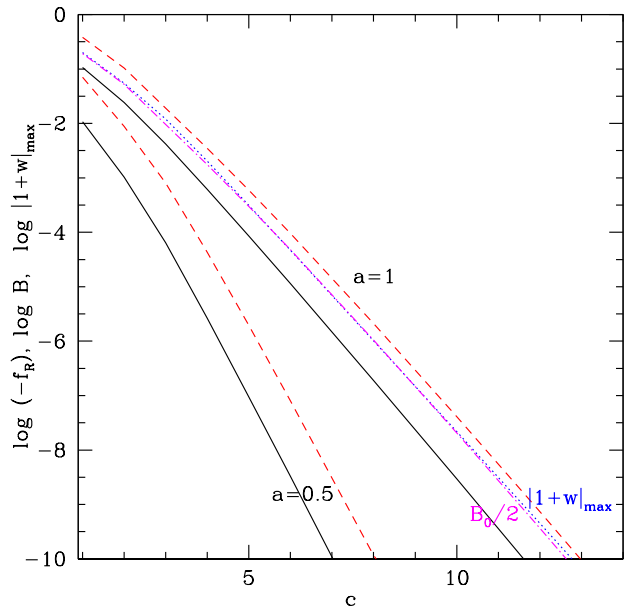,width=3.4in}
\caption{The derivative $f_R$ (solid black curves) and the structure 
parameter $B$ (dashed red), related to the Compton wavelength of the 
scalar part of the gravity modification, are plotted vs.\ $c$. 
The upper pair is evaluated at the present and the lower pair is at 
$a=0.5$.  
Note that $f_R$ (and $B$) has a very steep dependence on scale factor 
$a$, so structures that might nominally have $f_R\approx 4\times10^{-4}$ 
today may have 
formed when $f_R< 10^{-6}$ and so not exhibit observable deviations 
from general relativity.  The maximum deviation of the effective equation 
of state from a cosmological constant (dotted blue) closely follows 
$B(a=1)/2$ (dot-dash magenta). 
}
\label{fig:b0}
\end{center}
\end{figure}

Figure~\ref{fig:b0} shows the steep dependence of $f_R$ and $B$ with 
scale factor $a$ and how this allows a range for the characteristic 
scale $r$, or equivalently the parameter $c$ (related by 
Eq.~\ref{eq:dcr}).  For example, in the $n=4$ model of \cite{husaw} 
the value of $B$ drops by a factor 3 from the present to $a=0.5$ 
(redshift $z=1$), 
while the exponential model achieves a drop of 300 (30,000) for 
$c=5$ ($c=8$).  While the characteristic scale often seen in the 
literature is $m^2$, e.g.\ $f(R)$ varies as $(R/m^2)^n$, here the 
dependence $e^{-R/r}$ permits $r$ to lie in the range 
$r/m^2<3.8$ and still satisfy $f_R(a=0.5)<10^{-6}$.  Since $m^2$ 
is determined by the matter density today, this freedom somewhat eases 
the coincidence between the characteristic scale and today's curvature. 

This steep dependence also somewhat eases the condition for bound 
structures to recover general relativistic behavior: while the solar 
system, lying in the high curvature background of the galaxy, satisfies 
the conditions easily, our galaxy halo requires values of $f_R\lesssim 
10^{-6}$ (see \cite{husaw}, Sec.\ IIID for a detailed calculation). 
The sharp drop in $f_R$ when going to higher redshifts, as mentioned 
above, is stronger for the exponential model than most $f(R)$ theories. 
Still, galaxy constraints will play a significant role and we should 
be cautious of values of $c<4$ (imposing the condition at $a=0.5$, 
or $c<7$ if at 
$a=1$) failing to live up to the chameleon mechanism. 

In $f(R)$ models of modified gravity it is extremely difficult to 
get an appreciable deviation from $\Lambda$CDM behavior on the expansion 
side of observations while still obeying constraints from the structure 
side.  Using the very steep dependence of the modifications on curvature, 
and hence a rapid redshift dependence of the key structure parameter 
$f_R$ (or $B$), exponential gravity can again do better than most 
models to ameliorate the situation.  
If we were to impose $f_R<10^{-6}$, say, from structure constraints, this 
would imply a maximum deviation in the equation of state (occurring 
at the peak of the first phantom excursion) 
$|1+w|_{\rm max}<4\times 10^{-6}$ and $c>7$ if evaluated 
at the present, but $|1+w|_{\rm max}<2\times 10^{-3}$ and $c>4$ if 
evaluated at the time the structure forms, say $a=0.5$. 
So even pushing all values as far as allowed, it is almost impossible 
to get an appreciable signal in $1+w$.  
(Note that assuming the characteristic curvature modification scale is 
defined by the present matter density, i.e.\ $r/m^2=1$, gives 
$c=15.4$.)  Interestingly, we find $|1+w|_{\rm max}$ is excellently 
approximated by $(1/2)B(a=1)$, relating the expansion and structure 
sides.

\begin{figure}[!htb]
\begin{center}
\psfig{file=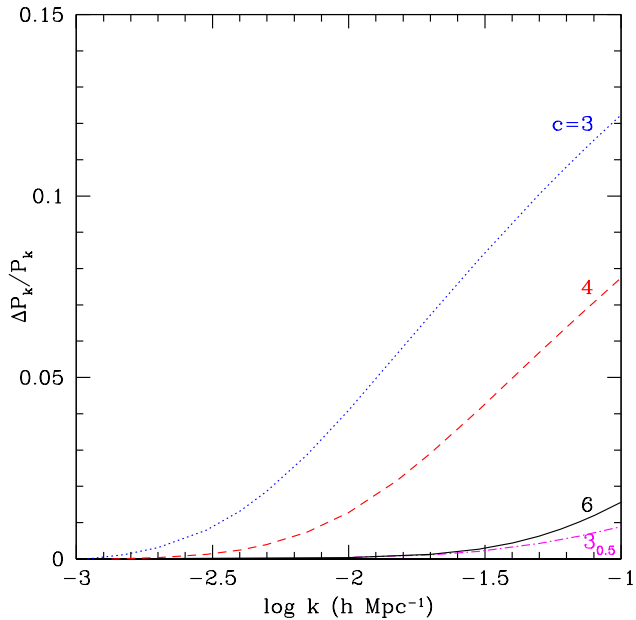,width=3.4in}
\caption{The matter power spectrum in $f(R)$ theory deviates from 
the $\Lambda$CDM result in a scale dependent manner.  
In the exponential gravity case, a wide range of 
values of $c$ have suppressed deviations and can be consistent with 
observations.  The lowest curve has the same value $c=3$ as the 
top curve, but plots the power spectrum deviation at $a=0.5$, 
showing the effect of the steepness of the model. 
}
\label{fig:grow}
\end{center}
\end{figure}

We can check the magnitude of deviations from $\Lambda$CDM growth of 
structure, at least in the linear regime, by solving for the 
evolution of linear density perturbations.  
We take into account both the change in the gravitational coupling 
strength (i.e.\ Newton's constant) and the scale-dependent effect of 
anisotropic stress in the $f(R)$ model (see, e.g., \cite{zhang,pogsil}). 
Figure~\ref{fig:grow} shows the fractional deviation in the matter power 
spectrum as a function of wavenumber today.

The deviations become noticeable on the $f_R$ Compton scale and have 
a magnitude determined by $B$.  Due to the steepness of the exponential 
model, the deviations are not as severe for a given $f_R(a=1)$ as 
for shallower $f(R)$ models.  In addition, the enhanced freedom in 
the characteristic scale allows models to be viable for values of $c$ 
much lower than the $c=15.4$ given by $r/m^2=1$.  
For $c>5$, the power spectrum deviations are less than 
5\% at $k=0.1\,h/$Mpc, beyond which linear treatment must give way 
to numerical simulations.  

Interestingly, the steepness also quickly 
reduces the deviations in the growth as one goes to higher redshift. 
The $c=3$ model with the largest deviations shown (12\% at the 
extreme), has less than 1\% deviations in the power spectrum at 
$a=0.5$.  Recall that $B$ can drop by two or three orders 
of magnitude in this model between $a=1$ and $a=0.5$. 
Note that because the expansion histories 
of $c>3$ models are so close to $\Lambda$CDM, one can interpret 
Fig.~\ref{fig:grow} also as showing the deviation in growth from 
a general relativity model with the same expansion history as the 
$f(R)$ models.  We have verified this directly by turning off the 
gravity modifications in the source term of the perturbation equation: 
the maximum change is below 0.2\%. 
Thus this figure shows the effects of modifying gravity for matched 
expansion.

\section{Conclusions \label{sec:concl}} 

The idea of explaining cosmic acceleration without adding a separate 
energy density component is attractive.  Einstein's general relativity 
has proved to be highly resistant to modification, however, and in 
excellent accord with observations.  Theories generalizing the linear 
factor of the Ricci scalar curvature to a more general function, 
$f(R)$ theories, are one of the main survivors for modifying 
gravity. 

In this paper we explored a simple model exponential in the 
curvature. 
This has several beneficial consequences, including just one more 
parameter than $\Lambda$CDM, steep dependence that enhances 
solar system and structure agreements, and amelioration of fine tuning 
of the characteristic curvature scale.  Most of the quantitative and 
qualitative results we present should be generally applicable to any 
sufficiently steep model however. 

The dynamics of the effective equation of state has several 
interesting properties, including ``sublimation'' rather than thawing 
from a cosmological constant state.  Unfortunately, even the loosened 
bounds on $f_R$ from structure constraints do not 
allow the equation of state to have detectable deviations from 
a cosmological constant expansion history -- one can view the failure 
for exponential models, steeper than any power law, as an indication of 
the difficulty in distinguishing viable $f(R)$ theories' expansion from 
the cosmological constant.  
We also find an excellent fitting form for the maximum 
deviation from $w=-1$ in terms of the Compton scale $B$, a direct 
relation between the expansion and structure characteristics. 

For structure growth, there exists an enlarged parameter space 
that gives observationally allowed deviations from $\Lambda$CDM 
in the matter power spectrum.  In particular, because of the steep 
evolutionary dependence, $f_R$ today may be as large as $\sim 10^{-3}$ 
and still agree with structure constraints.  The sort of model discussed 
here is of interest in that we have increased freedom to have a viable 
$f(R)$ modification of Einstein gravity 
that is yet distinct from standard $\Lambda$CDM.

\acknowledgments 

I thank Wayne Hu and Tristan Smith for useful discussions. 
This work has been supported in part by the Director, Office of Science, 
Office of High Energy Physics, of the U.S.\ Department of Energy under 
Contract No.\ DE-AC02-05CH11231, and World Class University grant 
R32-2008-000-10130-0 in Korea.


\begin{thebibliography}{99}

\bibitem{durrer} 
R. Durrer \& R. Maartens 2008, arXiv:0811.4132

\bibitem{araa} 
R.R. Caldwell \& M. Kamionkowski 2009, Ann. Rev. Astron. Astrophys. to 
appear [arXiv:0903.0866] 

\bibitem{sotiriou} 
T.P. Sotiriou \& V. Faraoni 2008, Rev. Mod. Phys. to appear 
[arXiv:0805.1726] 

\bibitem{songhusaw}
Y-S. Song, W. Hu, I. Sawicki 2007, Phys. Rev. D 75, 044004 
[arXiv:astro-ph/0610532] 

\bibitem{husaw} 
W. Hu \& I. Sawicki 2007, Phys. Rev. D 76, 064004 [arXiv:0705.1158]

\bibitem{chibasmith} 
T. Chiba, T.L. Smith, A.L. Erickcek 2007, Phys. Rev. D 75, 124014 
[arXiv:astro-ph/0611867] 

\bibitem{zhang}
P. Zhang 2006, Phys. Rev. D 73, 123504 [arXiv:astro-ph/0511218] 

\bibitem{appbat} 
S.A. Appleby \& R.A. Battye 2007, Phys. Lett. B 654, 7 [arXiv:0705.3199] 

\bibitem{tsuji}
S. Tsujikawa 2008, Phys. Rev. D 77, 023507 [arXiv:0709.1391] 

\bibitem{odintsov} 
G. Cognola et al. 2008, Phys. Rev. D 77, 046009 [arXiv:0712.4017] 

\bibitem{cahndl} 
R.N. Cahn, R. de Putter, E.V. Linder 2008, JCAP 0811, 015 [arXiv:0807.1346] 

\bibitem{caldlin} 
R.R. Caldwell \& E.V. Linder 2005, Phys. Rev. Lett. 95, 141301 
[arXiv:astro-ph/0505494] 

\bibitem{khoury} 
J. Khoury \& A. Weltman 2004, Phys. Rev. D 69, 044026 [arXiv:astro-ph/0309411] 

\bibitem{simul} 
H. Oyaizu, M. Lima, W. Hu, Phys. Rev. D 78, 123524 [arXiv:0807.2462] 

\bibitem{pogsil} 
L. Pogosian \& A. Silvestri 2008, Phys. Rev. D 77, 023503 [arXiv:0709.0296] 

\end{thebibliography}
\end{document}